\documentclass[11pt]{article}
\pdfoutput=1

\usepackage[no-natbib-sort]{jheppub}
\usepackage[T1]{fontenc} 

\usepackage{amssymb}
\usepackage{amsfonts}
\usepackage{amsbsy}
\usepackage{amsmath}
\usepackage{amsthm}
\usepackage{graphicx}
\usepackage{epstopdf}
\usepackage[vcentermath]{youngtab}
\usepackage{multirow}
\usepackage{latexsym} 
\usepackage{array}
\usepackage{color}
\usepackage{verbatim}
\usepackage[utf8]{inputenc}

\input cyracc.def 
\newfont{\twelvecyr}{wncyr10 at 12pt}

\def\Z{\mathbb{Z}}

\def\P{\mathbb{P}}

\def\n3a{t}

\newcommand{\SU}[0]{\mathrm{SU}}

\newcommand{\U}[0]{\mathrm{U}}

\newcommand*{\cy}{CY }

\newcommand{\kobe}[1]{\footnote{\textcolor{blue}{\textbf{KL:\ #1}}}}

\title{Large U(1) charges from flux breaking in 4D F-theory models}

\author{Shing Yan Li}
\author{and Washington Taylor}

\affiliation{Center for Theoretical Physics\\
Department of Physics\\
Massachusetts Institute of Technology\\
77 Massachusetts Avenue\\
Cambridge, MA 02139, USA}

\emailAdd{sykobeli at mit.edu}
\emailAdd{wati at mit.edu}

\preprint{MIT-CTP/5465}

\abstract{ We study the massless charged spectrum of $\U(1)$ gauge
  fields in F-theory that arise from flux breaking of a nonabelian
  group.  The $\U(1)$ charges that arise in this way can be very large.
  In particular, using vertical flux breaking, we construct an
  explicit 4D F-theory model with a $\U(1)$ decoupled from other gauge
  sectors, in which the massless/light fields have charges as large as
  657. This result greatly exceeds prior results in the literature. We
  argue heuristically that this result may provide an upper bound on
  charges for light fields under decoupled $\U(1)$ factors in the F-theory landscape. We
  also show that the charges can be even larger when the $\U(1)$ is
  coupled to other gauge groups.  }

\begin{document}
\maketitle
\flushbottom
\section{Introduction}
\label{sec:Intro}

It is well-known that string theory, when
compactified on manifolds in various dimensions,
gives a vast range of vacuum solutions known as the
string landscape. The low-energy physics of
these vacuum solutions can be
described by quantum field theories coupled to
gravity, with a wide range of
different gauge groups and matter
content. Nevertheless, there are strong constraints
from string theory, or quantum gravity in general,
on the low-energy theories that have a consistent
UV completion with gravity. 
Such constraints have been a key component in the analysis of string
theory since the early days of the subject, when Green and Schwarz
identified the strong conditions imposed by anomaly cancellation on
quantum theories of gravity in ten dimensions \cite{Green:1984sg},
leading to the identification of the heterotic string theory
\cite{Gross:1984dd}; later work has shown that indeed all consistent
theories of quantum gravity in ten dimensions with supersymmetry are
those that come from string theory (at least at the level of massless
spectra) \cite{Adams:2010zy,Kim:2019vuc}.

In lower space-time dimensions, particularly in 4D, the relationship
is much less clear
between the set of theories that can be realized in string theory and
those that appear consistent from the point of view of low-energy EFT
coupled to gravity.  In recent years, observations of general features
of string vacua and black hole behavior have led to a number of
speculations regarding quantum gravity constraints on low-energy
physics that are often referred to as  \emph{Swampland
  Conjectures} \cite{VafaSwamp,OoguriVafaSwamp} (see \cite{vanBeest:2021lhn} for a recent
review). These constraints have the potential
not only to shed light on
the general structure of string theory and quantum
gravity, but also may have phenomenological
implications leading to insights into physics
beyond the Standard Model.

One concrete set of questions regarding consistent quantum gravity
theories and the string landscape addresses bounds on the complexity
of the gauge and matter fields that are possible.
For example, while the rank of the gauge group in 6D or 4D
supersymmetric
string
vacua can be very high (see, e.g., \cite{Candelas:1997eh,Aspinwall:1997ye,MorrisonTaylorToric,Wang:2020gmi}), there is believed to be a
finite bound.  Similarly,
explicit string constructions of vacua in four and
higher dimensions give massless or light matter
representations of bounded complexity for nonabelian gauge groups
(see, e.g., \cite{Dienes:1996yh,KleversEtAlExotic,Cvetic:2018xaq}).
In this paper we focus on the question of what kinds of charges are
possible for massless or light fields charged under a U(1) gauge group
in  a 4D string vacuum constructed from F-theory  \cite{VafaF-theory,MorrisonVafaI,MorrisonVafaII}.

One of the most
widely accepted swampland-style conjectures is the
\emph{Completeness Hypothesis} \cite{Polchinski:1998rr,Banks:2010zn}, which states that
in a gauge theory coupled to gravity, all gauge
charges (consistent with charge quantization) must
be realized by some physical states. 
This conjecture has been proven in the context of quantum gravity in
AdS space with a holographic dual description \cite{Harlow:2018tng}.
Here the
physical states can be massless, or massive
including black holes. 
On the other hand, the situation is less well understood if we
consider
only massless or light\footnote{By massless or light fields in the F-theory context, we mean states coming from
branes wrapping cycles with vanishing volume. In 4D, these
include both chiral fields, which are truly massless, and
vector-like fields, which are kinematically massless but get
some light masses (relative to black hole masses) in the low-energy
theory from interactions in the
superpotential. We discuss both cases in our examples.}
fields. We may expect upper bounds on the gauge
charges that can be realized by massless fields in
the landscape, but it is not clear how large the
upper bounds are or whether the bounds even exist.
This is particularly unclear for $\U(1)$ charges
since, as explained below, it is very hard to
geometrically engineer $\U(1)$ gauge groups with
even moderately large charges (i.e., $q > 3$)
for massless states in string theory.

It is natural to look for such upper bounds using
the framework of F-theory, since this approach provides a
global description of the largest 
connected class of supersymmetric string vacua that is currently understood (see
\cite{WeigandTASI} for a review). 
F-theory gives 4D
$\mathcal N=1$ supergravity models when
compactified on elliptically fibered Calabi-Yau
(CY) fourfolds $Y$, corresponding to
non-perturbative compactifications of type IIB
string theory on general
(non-Ricci flat) complex K\"ahler threefold base
manifolds $B$. F-theory is also known to contain many vacua that are dual to many other types
of string compactifications, such as heterotic
models. The power of F-theory comes from
geometrizing the non-perturbative 7-brane
backgrounds in type IIB string theory into
elliptically fibered manifolds, which can be
analyzed using well-established tools in algebraic
geometry. Therefore, F-theory allows us to explore
the strongly coupled regime of the string landscape.
Charge completeness in the
context of F-theory is shown in \cite{Morrison:2021wuv} to follow from
some standard assumptions regarding the physical interpretation of the
F-theory geometry, for 6D theories and corresponding gauge sectors of
4D theories.

In the F-theory framework, nonabelian
gauge groups arise from singularities on divisors (algebraic
codimension-one loci) on $B$. In six or more space-time dimensions,
the form of the nonabelian part of the gauge group and corresponding
massless matter content can be easily determined using the local
geometry \cite{Kodaira,Neron,BershadskyEtAlSingularities,KatzVafa},
which is easy to study.  In contrast, $\U(1)$ gauge factors in 6D and 8D
F-theory models, as well as in many 4D models, arise from the global geometry. To
be precise, these abelian factors in the gauge group arise
 from a Mordell-Weil group of rational sections
with nonzero rank in the elliptic fibration
\cite{MorrisonVafaII,AspinwallMorrisonNonsimply,Aspinwall:2000kf,Grimm:2010ez}. It
is much harder to engineer these models, and surprisingly few explicit
F-theory constructions have been found with any but the simplest
charged matter structure. The best-understood class of models with
a single $\U(1)$, known as the Morrison-Park model
\cite{MorrisonParkU1}, 
gives a universal form of Weierstrass model with U(1) gauge group and
 massless (absolute values of)
charges\footnote{Throughout the paper, we normalize the nonzero
  charges such that they are all integers with the greatest common
  divisor being 1.} $q=1,2$.  Explicit models with $q=3,4,5$
have been constructed in \cite{KleversEtAlToric,Raghuram34,Knapp:2021vkm}
respectively, while models with $q=6$ are inferred from the type IIB
limit in \cite{Cianci:2018vwv}, and a procedure for constructing these
charges explicitly from universal flops is given in
\cite{Collinucci:2019fnh}. It has also been argued that $q$ can be
as large as 21 in 6D F-theory models using implicit Higgsing arguments
\cite{Raghuram:2018hjn}, 
and an algorithm for computing general U(1) charges from the form of a given
Weierstrass model has been developed in
\cite{Raghuram:2021wvx},
but explicit models with $q>6$ are still
  lacking. 
On the other hand, it was argued in
  \cite{Taylor:2018khc} that there is an infinite swampland of massless $\U(1)$
  charge spectra  in 6D supergravity theories.
In \cite{Raghuram:2020vxm}, a systematic criterion was proposed for
ruling out most of this infinite swampland, as F-theory constructions
of these models generally lead to an ``automatic enhancement'' of the
gauge group, and some low-energy arguments for this automatic enhancement
were put forth in \cite{Cvetic:2021vsw}.

Note that we primarily focus in this paper on charges of massless or light
fields under isolated $\U(1)$ factors only; more complicated charge
structures can arise when there are also nonabelian gauge factors
and there are fields that  have both $\U(1)$ and nonabelian charges,
as discussed in Section \ref{sec:coupling}.

The preceding discussion has focused primarily on 6D F-theory
models.
While 4D F-theory models  can be constructed with similar charges
using the same kinds of Weierstrass models described above
(Morrison-Park, etc.\ for charges up to $q = 6$), there
are also some qualitatively
different possibilities in 4D due to the inclusion of flux backgrounds,
which can affect the gauge groups and matter content.
In particular, with the power of fluxes it becomes
possible to build $\U(1)$ gauge groups from the
local geometry, which enables us to construct a much
larger class of $\U(1)$ models with larger $q$.
Indeed, it was noticed in \cite{LiFluxbreaking} that large $\U(1)$
charges can easily arise through breaking of
nonabelian gauge groups using  so-called
\emph{vertical} flux (referred to as  ``vertical flux
breaking''  henceforth), which will be described
below. In this paper, we take this approach.  We describe the general
framework of F-theory models with $\U(1)$ factors from flux breaking,
construct some examples with large charges ($q\gg 6$), and try
to identify a plausible upper bound for $q$ in the
4D F-theory landscape.

The strategy is as follows: We first identify
nonabelian models that support vertical flux
breaking down to a single decoupled $\U(1)$. We can
choose an arbitrarily exotic linear combination of
the Cartan $\U(1)$'s to be preserved, as long as
appropriate vertical flux satisfying all relevant
constraints is turned on. This exotic
$\U(1)$ is the source of large $q$.
As the combination becomes more exotic,
more flux is needed to satisfy flux quantization \cite{Witten:1996md},
and the flux configuration finally hits the tadpole
bound \cite{Sethi:1996es}. These are the only constraints that lead to
an upper bound of $q$ for a given geometry. 
We describe the general framework for this flux breaking and analyze
some specific models that give particularly large values of $q$.
To
maximize $q$, we should maximize the tadpole bound,
which is fixed by the Euler characteristic
$\chi(\hat Y)$ of the resolved elliptic Calabi-Yau fourfold $\hat Y$
from the F-theory construction.
At the same time, the general structure of the intersection form on
middle cohomology indicates that
we should minimize the
intersection numbers on the divisor $\Sigma$
that supports the original nonabelian factor, such that the
tadpole caused by a given flux configuration is
minimal. As a specific example of the  exotic $\U(1)$
charges arise from flux breaking, we construct an explicit 4D F-theory
model that  combines   the two optimizations described above,
leading to a surprisingly large value of $q$:
\begin{equation}
    q_\mathrm{max}= 657\,,
\end{equation}
for light vector-like charged matter fields.  A similar construction
can give truly massless  chiral matter fields with charges of 465 or
greater.

This paper is organized as follows: In Section
\ref{sec:fluxbreaking}, we review vertical flux and
the formalism of vertical flux breaking. The review
is brief, only presenting essential facts for
our constructions of $\U(1)$ models. We refer the
readers to \cite{LiFluxbreaking} for more details. 
In Section \ref{sec:general-formalism}, we go through the general
framework of vertical flux breaking from a geometric nonabelian group
to an isolated $\U(1)$ gauge factor, and illustrate with a specific
class of simple examples from the breaking $\SU(3) \rightarrow\U(1)$.
In Section
\ref{sec:u1}, we present the  explicit 4D F-theory model with
$q_\mathrm{max}=657$ for vector-like matter, and related models with
comparably large charges for massless chiral fields. The $\U(1)$ model comes from
a $G_2\rightarrow\U(1)$ breaking on the \cy
fourfold with the fifth highest $h^{3,1}$ in the
Kreuzer-Skarke (KS) database of toric hypersurface
constructions
\cite{Kreuzer:1997zg,Scholler:2018apc}. We describe this model in some
detail, and give qualitative arguments  for
why this model may give, or at least be close to, the upper
bound on decoupled $\U(1)$ charges in the 4D
F-theory landscape. In Section \ref{sec:coupling},
we extend our discussion to the case of $\U(1)$
coupled to other gauge groups, with an example of
even slightly larger $q_\mathrm{max}$ when the
$\U(1)$ is coupled to an $E_6$. We finally
conclude in Section
\ref{sec:Con}, and give more geometric properties
of our $\U(1)$ model in Appendix \ref{sec:equivalence}.

\section{Formalism of vertical flux breaking}
\label{sec:fluxbreaking}

In this section, we review the formalism of
breaking nonabelian gauge groups on divisors using
vertical flux in 4D F-theory models. As the
formalism has been described in depth in \cite{LiFluxbreaking}, here
we only recap the essential facts for our
construction of $\U(1)$ models and set up the notation.

\subsection{Vertical flux}
\label{subsec:vertical}
To describe the flux backgrounds, we first need some basic geometric
facts about the compactifications. As mentioned in Section
\ref{sec:Intro}, we consider F-theory compactified on a \cy fourfold
$Y$, which is an elliptic fibration on a threefold base
$B$. Nonabelian gauge groups arise when sufficiently high degrees of
singularities are developed in the elliptic fibers over divisors on $B$ (denoted by $D_\alpha$), called gauge
divisors $\Sigma$. When this happens, $Y$ itself is also singular and
we need to consider its resolution $\hat Y$ such that we can study
cohomology and intersection theory. Let the total gauge
group be $G$, where $G$ has no
$\U(1)$ factors before flux breaking. 
For clarity of the analysis, in this section we assume that $G$ is a
simple nonabelian gauge group, although essentially the same analysis
goes through when $G$ has multiple nonabelian factors, as in the cases
considered in \S\ref{sec:u1}.
The nonabelian group
 $G$ is supported on a gauge divisor $\Sigma$, and
the resolution
results in exceptional divisors $D_{1\leq{i}\leq\mathrm{rank}(G)}$
in $\hat Y$.  Their intersection structure matches
(up to monodromy for non-simply-laced groups) the Dynkin
diagram of $G$, where each exceptional divisor corresponds to a
Dynkin node \cite{Kodaira,Neron}. By the Shioda-Tate-Wazir theorem
\cite{shioda1972,Wazir}, the divisors $D_I$ on $\hat Y$ are spanned by
the zero section $D_0$ of the elliptic fibration, pullbacks of base
divisors $\pi^* D_\alpha$ (which we also call $D_\alpha$ depending on
context), and the exceptional divisors $D_{i}$.\footnote{If $G$ has
  $\U(1)$ factors, there are also divisors associated with these
  factors coming from the Mordell-Weil group of rational sections with
  nonzero rank.} Although the choice of resolution is not unique, our
analysis and results are clearly resolution-independent
\cite{Jefferson:2021bid}.

Now we are ready to understand fluxes. These are
most easily understood by considering the dual
M-theory picture of the F-theory models, that is,
M-theory compactified on the resolved fourfold
$\hat Y$ (reviewed in \cite{WeigandTASI}). In the M-theory perspective, fluxes are
characterized by the three-form potential $C_3$ and
its field strength $G_4=dC_3$. The data of $G_4$
flux, which can be studied with well-established
tools, is sufficient for determining the gauge
groups with flux breaking.

In general, $G_4$ is a discrete flux that takes
values in the fourth cohomology $H^4(\hat Y,\mathbb R)$. Its quantization condition is given by \cite{Witten:1996md}
\begin{equation} \label{eq:fluxquantization}
    G_4+\frac{1}{2}c_2(\hat Y)\in H^4(\hat Y,\mathbb Z)\,,
\end{equation}
where $c_2(\hat Y)$ is the second Chern class of
$\hat Y$. In all the models we consider below,
the relevant components in $c_2$ are even and we just require that the
corresponding components in $G_4$  are integer quantized.

Next, to preserve the minimal amount of
supersymmetry (SUSY) and stability in 4D, $G_4$ must 
live in the $(2, 2)$ part of middle cohomology, i.e.,
 $G_4\in H^{2,2}(\hat Y,\mathbb R)\cap H^4(\hat Y,\mathbb Z)$. SUSY also
imposes the condition of primitivity \cite{Becker:1996gj,Gukov:1999ya}:
\begin{equation} \label{eq:primitivity}
    J\wedge G_4=0\,,
\end{equation}
where $J$ is the K\"ahler form of $\hat Y$.
Typically primitivity is automatically satisfied,
but this is not the case when there is vertical flux
breaking. In our models, the primitivity condition leads to
stabilization of some K\"ahler moduli, and
stabilization within the K\"ahler cone imposes constraints on $G_4$.

We also have the condition of D3-tadpole
cancellation for a consistent vacuum \cite{Sethi:1996es}:
\begin{equation} \label{eq:tadpole}
    \frac{\chi(\hat Y)}{24}-\frac{1}{2}\int_{\hat Y}G_4\wedge G_4=N_{D3}\in\mathbb Z_{\geq 0}\,,
\end{equation}
where $\chi(\hat Y)$ is the Euler characteristic of
$\hat Y$, and $N_{D3}$ is the number of D3-branes.
To preserve SUSY and stability, we require that
there are no anti-D3-branes i.e. $N_{D3}\geq 0$.
This condition constrains the size of fluxes to a
finite number, which, as shown in the next section,
also limits the size of $\U(1)$ charges that can be realized.

All the above constraints are satisfied by general
fluxes, while the flux breaking in this paper only
uses \emph{vertical} flux, which satisfies extra
constraints. To study this, first consider the
orthogonal decomposition of the middle cohomology \cite{Braun:2014xka}:
\begin{equation}
    H^{4}(\hat Y,\mathbb C)=H^{4}_\mathrm{hor}(\hat Y,\mathbb C)\oplus H^{2,2}_\mathrm{vert}(\hat Y,\mathbb C)\oplus H^{2,2}_\mathrm{rem}(\hat Y,\mathbb C)\,.
\end{equation}
Here the summands refer to horizontal, vertical,
and remainder fluxes respectively. The vertical
subspace is spanned by products of harmonic
$(1,1)$-forms (which are Poincar\'e dual to
divisors, denoted by $[D_I]$)
\begin{equation}
    H^{2,2}_\mathrm{vert}(\hat Y,\mathbb C)=\mathrm{span}\left(
    H^{1,1}(\hat Y,\mathbb C)\wedge H^{1,1}(\hat Y,\mathbb C)\right)\,.
\label{eq:vertical-c}
\end{equation}
According to Eq.\ (\ref{eq:fluxquantization}),
vertical flux should live in the integral vertical
subspace $H^{2,2}_\mathrm{vert}(\hat Y,\mathbb
R)\cap H^4(\hat Y,\mathbb Z)$ (when $c_2$ is even). This subspace is in
general hard to analyze, hence we only focus on a
slightly smaller subspace $H^{2,2}_\mathrm{vert}(\hat Y,\mathbb Z)$, which is
defined as
\begin{equation}
    H^{2,2}_\mathrm{vert}(\hat Y,\mathbb Z):=\mathrm{span}_{\mathbb Z}\left(
    H^{1,1}(\hat Y,\mathbb Z)\wedge H^{1,1}(\hat Y,\mathbb
    Z)\right)\,.
\label{eq:vertical-z}
\end{equation}
That is, the span of integer multiples of forms
$[D_{I}] \wedge[D_J]$. This subspace, although it may
be smaller than the full integral vertical subspace
in general, provides the structure we need for interesting
phenomena from flux breaking. We leave the full
analysis of integral vertical flux to future work.

Here are some notations for analyzing vertical flux. We expand
\begin{equation}
    G_4^\mathrm{vert}=\phi_{IJ} [D_I]\wedge [D_J]\,,
\label{eq:g-phi}
\end{equation}
and work with integer flux parameters $\phi_{IJ}$.
We will specify the basis for the expansion later.
Next, we denote the integrated flux as \cite{Grimm:2011fx}
\begin{equation}
    \Theta_{\Lambda\Gamma}=\int_{\hat Y} G_4\wedge [\Lambda]\wedge
          [\Gamma]\,,
\label{eq:integrated-flux}
\end{equation}
where $\Lambda,\Gamma$ are arbitrary linear
combinations of $D_I$, and subscripts $0, i, \alpha$ refer to the
basis divisors $D_0, D_{i}, D_\alpha$. In this paper, we use the
following
resolution-independent
 formula to relate
$\Theta_{i \alpha}$ to $\phi_{i \alpha}$
\cite{Grimm:2011sk,Jefferson:2021bid}:\footnote{Indices appearing twice are
summed over, while other
  summations are indicated explicitly.}
\begin{equation} \label{eq:phitoTheta}
    \Theta_{i \alpha}=-\kappa^{i j} \Sigma\cdot D_\alpha\cdot D_\beta \phi_{j\beta}\,,
\end{equation}
where $\kappa^{i j}$ is the inverse Killing
metric of $G$, and ``dots'' denote the
intersection product.\footnote{We will not mention
explicitly the space where
the products are taken in such formulae, as the space ($\hat{Y}$ or
$B$, the latter in this case)
is already clear from context.}
This is the same as
the Cartan matrix of $G$ for ADE groups, but in general it is not.
Note that in our examples, the only nonzero flux parameters
 have indices of type $\phi_{j \alpha}$. In general, many gauge groups
 also admit fluxes of type $\phi_{ij}$ associated with chiral matter
 of the nonabelian gauge group.  The integrated
fluxes $\Theta_{j\alpha}$ can also be affected by these parameters,
in a fashion that also seems to have a resolution-independent
description \cite{Jefferson:2021bid}, although we will not use such
fluxes here.

Now we write down the extra flux constraints
satisfied by vertical flux. First, to preserve
Poincar\'e symmetry after dualizing from M-theory, we require \cite{Dasgupta:1999ss}
\begin{equation} \label{eq:PoincareSym}
    \Theta_{0\alpha}=\Theta_{\alpha\beta}=0\,.
\end{equation}
If the whole $G$ is preserved, a necessary condition is that
\begin{equation} \label{eq:gaugeSym}
    \Theta_{i\alpha}=0\,,
\end{equation}
for all $i,\alpha$. This condition is also
sufficient when there is no nontrivial remainder
flux. Vertical flux breaking occurs when this
condition is violated, which we will discuss next.
Note that the violation does not affect the
condition in Eq.\ (\ref{eq:PoincareSym}).

\subsection{Vertical flux breaking}
\label{subsec:breaking}
With the knowledge of vertical flux, we now
describe the breaking of geometric gauge groups
with vertical flux, or vertical flux breaking. This
kind of breaking has been used as early as \cite{BeasleyHeckmanVafaI} (see also \cite{WeigandTASI}), and
is recently developed in depth in \cite{LiFluxbreaking}. In this
paper, we only list the results essential for our
analysis on $\U(1)$ models, and we refer readers to
\cite{LiFluxbreaking} for full technical details. Note that flux
breaking can also be done with remainder flux \cite{Buican:2006sn,BeasleyHeckmanVafaII}, but
as noted in \cite{LiFluxbreaking}, vertical flux should be used to
realize exotic $\U(1)$ charges.

Recall that we need $\Theta_{i \alpha}=0$ for all
$i,\alpha$ to preserve the whole $G$. Now we
break $G$ into a smaller group $G'$ by turning on
some nonzero $\phi_{i\alpha}$. Such flux breaks
some of the roots in $G$. It also induces masses
for some Cartan gauge bosons by the St\"uckelberg
mechanism \cite{Grimm:2010ks,Grimm:2011tb}, hence breaks some combinations of Cartan
$\U(1)$'s in $G$. 
Let $\alpha_{i}$ be the simple
roots of $G$, and $T_{i}$ be the Cartan generators
associated with $\alpha_{i}$ i.e. in the co-root
basis. The root $b_{i} \alpha_{i}$ is preserved under the breaking if
\begin{equation} \label{eq:fluxbreakingcondition}
    \sum_{i} b_{i}\left<\alpha_{i},\alpha_{i}\right>\Theta_{i \alpha}=0\,, 
\end{equation}
for all $\alpha$. Here $\left<.,\,.\right>$ denotes
the inner product of root vectors. Moreover, the
corresponding linear combination of Cartan generators
\begin{equation}
    \sum_{i} b_{i}\left<\alpha_{i},\alpha_{i}\right>T_{i}\,, 
\end{equation}
is preserved. These generators form a nonabelian
gauge group $G'\subset G$ after breaking.

There are additional constraints on vertical flux
breaking coming from primitivity, since Eq.\ 
(\ref{eq:primitivity}) is not automatically
satisfied when there is vertical flux breaking and
$\Theta_{i\alpha}\neq 0$ for some $i,\alpha$.
In particular, primitivity requires that
\begin{equation} \label{eq:intprimitivity}
    \int_{\hat Y}[D_{i}]\wedge J\wedge G_4=0\,,
\end{equation}
which is true only for specific choices of $J$ when
there is vertical flux breaking. As a result, the condition of
primitivity stabilizes some but not all K\"ahler
moduli in $J$. 
As discussed in \cite{LiFluxbreaking}, 
 in the presence of flux breaking, there is a nontrivial condition on
 the $\alpha$ components of $J$
(in an expansion  $J = t^I [D_I]$ in $\hat{Y}$)
that must be
satisfied to ensure a nontrivial solution of
K\"ahler moduli, which can be described as follows:
Let $r$ be the number of
linearly independent $D_\alpha$'s appearing
 in the set of all homologically
independent surfaces in the form of $S_{i \alpha}=D_{i}\cdot
D_\alpha$
(for any $i$ of the given $G$). Now consider the ($r$  $\times$
$\mathrm{rank}(G)$)
matrix $\Theta_{(\alpha_{a})(i)}$ (where $a$ and $i$
are the indices for rows and columns respectively). 
The condition (\ref{eq:intprimitivity}) asserts that $t^\alpha
\Theta_{\alpha i} = 0$.
Since the solution
to primitivity thus requires a nontrivial left null space of the matrix,
the rank of the matrix is at most $r-1$. Moreover from Eq.
(\ref{eq:fluxbreakingcondition}), the rank of the
matrix is also the change in rank of $G$ during flux breaking. Therefore, we require
\begin{equation} \label{eq:rs}
    r\geq\mathrm{rank}(G)-\mathrm{rank}(G')+1\,.
\end{equation}
In particular, when remainder flux breaking is not
available, 
and all divisors in $\Sigma$ descend
from intersections in $B$, we have
$r=h^{1,1}(\Sigma)$. This condition limits the
availability of vertical flux breaking, which plays
an important role in the analysis below. 
There are
still additional sign constraints on the fluxes in
order to stabilize the K\"ahler moduli within the
K\"ahler cone. These constraints will be explicitly demonstrated
in examples below.

The above vertical flux generically also induces chiral matter
charged under $G'$ if $G'$ (regardless of $G$) supports
chiral matter. In this paper, we focus on
cases where matter is charged under a single simple
nonabelian gauge group $G$, and
 $G$ does
not support chiral matter.
Then the chiral indices are given by the
following: for a weight $\beta=-b_{i}\alpha_{i}$ in a representation
$R$ of $G$ that is localized on the matter curve $C_R=\Sigma\cdot D_R$,
by analysis following
\cite{Braun_2012,Marsano_2011,KRAUSE20121,Grimm:2011fx} the chiral
index is
\begin{equation} \label{eq:chi}
    \chi_\beta=\sum_{i}b_{i}\frac{\left<\alpha_{i},\alpha_{i}\right>}{\left<\alpha_{\mathrm{max}},\alpha_{\mathrm{max}}\right>}\Theta_{i D_R}\,,
\end{equation}
where $\alpha_{\mathrm{max}}$ is the longest $\alpha_{i}$. If $R$ is
the adjoint localized on the bulk of $\Sigma$, we should
replace $C_R$ by the canonical class $K_{\Sigma}$ i.e.
$D_R=K_B+\Sigma$ by adjunction.

In addition to nonabelian gauge factors, flux breaking can also give
rise to $U(1)$ gauge factors, either in combination with nonabelian
factors as studied in \cite{LiFluxbreaking}, or in isolation. The
latter situation is the main focus of this work.

\section{Flux breaking to $\U(1)$}
\label{sec:general-formalism}

We now turn to abelian $\U(1)$ factors in $G'$,
which are a key feature of vertical flux breaking.
We start by giving the general framework for flux breaking  of a
simple nonabelian factor to $\U(1)$ and then give a simple
illustrative example of breaking $\SU(3) \rightarrow\U(1)$ using
vertical fluxes.

\subsection{$\U(1)$ factors from flux breaking}

Although every root of a simple Lie algebra corresponds to a linear
combination of Cartan generators, the reverse is
seldom true. In fact, we can write down arbitrary
linear combinations of Cartan generators, while
there is only a finite number of roots. Following
the logic of vertical flux breaking described in
\S\ref{subsec:breaking}, 
suppose that we have an F-theory model over a threefold base $B$ that
contains a single nonabelian gauge factor $G$.
We then turn on vertical flux parameters $\phi_{j\beta}$ giving
some non-vanishing fluxes $\Theta_{i\alpha}$ through Eq.\ (\ref{eq:phitoTheta}).
If we impose the condition (summing as above by convention over
doubled indices $i$)
\begin{equation} \label{eq:exoticu1}
    p_{i}\Theta_{i\alpha}=0\,,
\end{equation}
for all $\alpha$, while
\begin{equation}
    \sum_{i} p_{i}\left<\alpha_{i},\alpha_{i}\right>^{-1}\alpha_{i}\,,
\end{equation} is not along any roots, then the Cartan generator
\begin{equation}
    p_{i}T_{i}\,,
\end{equation}
is preserved but does not belong to any nonabelian subgroup of
$G$. 
Such generators thus form
the abelian part of the preserved gauge group $G'$. More generally, there may be $\U(1)$'s
that are combinations of Cartan generators from
multiple gauge factors. These $\U(1)$'s, however, are
not relevant to our analysis below, and we focus on
$\U(1)$'s coming from  $G$ with a single simple nonabelian factor as above.
We focus attention in particular on cases
involving vertical flux breaking of
such a gauge group
$G$,  where no nonabelian gauge factor
remains and we have a single residual $\U(1)$ gauge
factor on the gauge divisor.
 As
long as Eq.\ (\ref{eq:exoticu1})
is satisfied, the
coefficients $p_{i}$ look arbitrary and the
resulting $\U(1)$ can naively be arbitrarily complicated, which leads
to arbitrarily exotic matter, although as we shall discuss there are
upper bounds from other flux constraints. This
feature gives great power for building $\U(1)$
models from vertical flux breaking, as one can flexibly tune
suitable $p_{i}$ to get a desired $\U(1)$, with specific matter content. In
contrast, in field theory for example, the $\U(1)$ realized
after breaking through a Higgsing process is determined by the representation
and vev of the Higgs field, which substantially constrains the
resulting possible $\U(1)$ factors and associated charges.  While these
kinds of Higgsed $\U(1)$ fields are transparent from the low-energy physics
point of view, they are
much harder
to study in general in F-theory as they involve deformations of the
Weierstrass model that are in some cases unknown
\cite{Raghuram:2018hjn}. In contrast, vertical flux
breaking seems to rely much on the UV
physics of string theory, and while we have a clear way of analyzing these
systems from the geometry of fluxes,
so far we do not see any
clear approach to attaining a low-energy description of the
breaking.

The condition in Eq.\ (\ref{eq:rs}) also holds for $\U(1)$ factors.
Now $\mathrm{rank}(G')$ also counts the number of $\U(1)$'s that
descend from $G$. In particular, to break a high-rank nonabelian
factor to a single $\U(1)$, the
gauge factor must arise on a divisor with $h^{1, 1} (\Sigma)\geq
\mathrm{rank} (G)$.

\subsection{A simple example: breaking $\SU(3) \rightarrow\U(1)$}
\label{subsec:simpleexample}

It is useful to demonstrate the above techniques with a simple example of
$\U(1)$ models before discussing the maximization of $\U(1)$ charges.
Let us consider the base $B$ as a $\mathbb P^1$-bundle over $\mathbb F_0=\mathbb P^1\times\mathbb P^1$, with an $\SU(3)$ supported on
$\mathbb F_0$.\footnote{The analysis here is independent of how the
$\SU(3)$ is realized on $\mathbb F_0$.} Notice that $\SU(3)$ has rank
$2$ and $h^{1,1}(\mathbb F_0)=2$. Therefore by Eq.\ (\ref{eq:rs}), $B$
is the simplest base that supports the breaking to $\U(1)$
described in the last subsection. Since the models in the coming sections
have the same divisor geometry, this subsection also serves as a
warm-up exercise for those constructions.

First, we describe the geometry of $B$. Let the two
$\mathbb P^1$'s on $\mathbb F_0$ be $s,f$. Then $B$ has three
independent divisors: $\Sigma$ as the section $\mathbb F_0$,
and $S,F$ as the $\mathbb P^1$ bundles on $s,f$ respectively. $\Sigma$
is also the gauge divisor. The only nontrivial intersection number is $\Sigma\cdot S\cdot F=1$. Generically
there are (anti-)fundamentals $\mathbf 3$ and $\bar{\mathbf 3}$, as
well as the adjoint $\mathbf 8$ as the matter content of the model.

Eq.\ (\ref{eq:rs}) tells us that we can at most reduce the rank of
the gauge group by one when satisfying primitivity. In other words,
to preserve at least a $\U(1)$,
the nonzero vertical flux should always satisfy the constraint
\begin{equation}
    a\Theta_{1\alpha}+b\Theta_{2\alpha}=0\,,
\end{equation}
for all $\alpha=S,F$, where $1,2$ are the Cartan indices for $\SU(3)$,
and $a,b$ are some coprime integers. There are two possible cases: for
generic $a,b$ we obtain the breaking $\SU(3)\rightarrow\U(1)$, but if
$a=0$, $b=0$, or $a=b$, these coefficients align with some roots of
$\SU(3)$ and we get $\SU(3)\rightarrow\SU(2)$ instead. We focus on
the former case with the $\U(1)$ generator $T=aT_1+bT_2$. The flux
constraint is then solved by
\begin{equation}
    \left(\phi_{1S},\phi_{2S},\phi_{1F},\phi_{2F}\right)=\frac{1}{3}\left(\left(a-2b\right)n_{S},\left(2a-b\right)n_{S},\left(a-2b\right)n_{F},\left(2a-b\right)n_{F}\right)\,,
\end{equation}
where $n_S,n_F$ are flux parameters to be chosen, such that all
$\phi$'s are integers to satisfy flux quantization.

Now we turn to the condition of primitivity. In the
F-theory limit where the elliptic and exceptional
fibers shrink to zero volume, only the K\"ahler
form of $\Sigma$ contributes in Eq.\ 
(\ref{eq:intprimitivity}). Let the
K\"ahler form of $\Sigma$ be
\begin{equation}
    [J_\Sigma]=t_1 \Sigma\cdot F+t_2\Sigma\cdot S\,,
\end{equation}
where $t_1,t_2$ are K\"ahler moduli. Eq.\ (\ref{eq:intprimitivity}) then implies
\begin{equation}
    t_1 n_S+t_2 n_F=0\,.
\end{equation}
To ensure stabilization within the K\"ahler cone
where $t_1,t_2>0$, we require $n_S,n_F$ to be both
nonzero and have opposite signs.

Assuming the tadpole constraint 
(\ref{eq:tadpole}) is satisfied,
now we are free to choose the parameters $a,b,n_S,n_F$ and
calculate the resulting $\U(1)$ charges. First notice that under
the breaking, the (anti-)fundamentals give charges $a,b,a-b$ and
their conjugates, and the adjoint gives charges $a+b,a-2b,2a-b$ and
their conjugates. As an example with small flux parameters, let us
choose $(a,b,n_S,n_F)=(-2,5,2,-1)$. Then we obtain the following spectrum:
\begin{equation}
    q=2,\,3,\,5,\,7,\,9,\,12\,.
\end{equation}
To be more precise, we can also calculate the chiral spectrum of
these charges. Using Eq.\ (\ref{eq:chi}), we see that the chiral
spectrum induced from the adjoint is
\begin{equation}
    14\times\mathbf 1_3+4\times\mathbf 1_{12}+10\times\mathbf 1_{-9}\,.
\end{equation}
It is easy to check that this chiral spectrum is free of both pure
gauge and gauge-gravity anomalies
since $\sum q_i =\sum q_i^3 =  0$. More generally, for any
such $a,b$, we have the following chiral spectrum from the adjoint:
\begin{equation}
    2(b-a)\times\mathbf 1_{a+b}+2a\times\mathbf
    1_{a-2b}+2b\times\mathbf 1_{2a-b}\,,
\label{eq:infinite-family}
\end{equation}
which is remarkably always anomaly-free.
One can perform a similar analysis for the (anti-)fundamentals,
although it depends more on the geometry of $B$.

Notice that the charges are coprime. Therefore through such a
simple construction, we already obtain some relatively large
$\U(1)$ charges. In the next Section, we will optimize this
procedure subject to the tadpole constraint, to obtain our
extremal result $q_\mathrm{max}=657$.

Some relevant comments can be made
here regarding the connection of these spectra with related 6D models.
The family of models described in Eq.\ (\ref{eq:infinite-family}) is
very similar in structure to an infinite family of 6D U(1) models with
arbitrarily large charges encountered in \cite{Taylor:2018khc}.  In
the 6D case, charges arise from complicated Weierstrass models (see,
e.g., \cite{Raghuram:2021wvx}), and the infinite family is apparently
rendered unphysical by the automatic enhancement mechanism
\cite{Raghuram:2020vxm,Cvetic:2021vsw}, which guarantees the
appearance of an additional U(1) factor.  In the 4D case, the charges
arise from the distinct physical mechanism of flux breaking, so the
infinite family of anomaly-consistent models is bounded by the
tadpole, and automatic enhancement does not seem to occur.  It would
be interesting to better understand how the automatic enhancement
story differs in this context.  It is also interesting to observe that
because $a, b$ are coprime, this family of models can
 contain massless
or light matter fields that generate the full charge lattice, in
accord with the massless charge sufficiency conjecture formulated for
6D F-theory in \cite{Morrison:2021wuv}.  In this case, however, the
nonzero multiplicities of massless or light matter depend upon the
choice of flux.  It would be interesting to look further into the
question of whether the light fields always
generate the full charge lattice
for arbitrary choices of flux.

\section{A $\U(1)$ model with $q_\mathrm{max}=657$}
\label{sec:u1}
In this section, we construct a $\U(1)$ model with
$q_\mathrm{max}=657$ using vertical flux breaking.
We describe the geometry of the fourfold $\hat Y$
and the base $B$, as well as the vertical flux
background in detail. Then we give qualitative and
heuristic arguments towards $q_\mathrm{max}=657$
being (close to) an upper bound in the 4D F-theory landscape.
Notice that there are other nonabelian gauge
factors in this model, but they are completely
decoupled from the $\U(1)$ we construct, hence we
still call it a $\U(1)$ model, and the analysis of
\S\ref{sec:general-formalism} applies essentially unchanged.

It is useful to first recap our strategy. From Eq.\ 
(\ref{eq:exoticu1}), we see that the more exotic
the $\U(1)$ or the coefficient $p_{i}$ is, the
larger integer $\phi_{i\alpha}$ we need to turn
on. From Eq.\ (\ref{eq:tadpole}), the size of
$\phi_{i\alpha}$ is bounded from above by the
Euler characteristic $\chi(\hat Y)$ and the
intersection numbers on $\Sigma$ that arise in
$[G_4]\cdot[G_4]$. Therefore to obtain the largest
$q_\mathrm{max}$, we shall maximize $\chi(\hat Y)$
while minimizing the intersection numbers on
$\Sigma$. Although the list of elliptic \cy
fourfolds is far from complete, the KS database
provides a set of good toric representatives especially
at large $\chi$. Scanning through the KS database
leads us to consider the \cy fourfold with the
fifth largest $h^{3,1}$ and $\chi$. We now describe its geometry in detail.

\subsection{Geometry}
\label{subsec:geometry}
The fourfold $\hat Y$ has the following Hodge numbers:
\begin{gather} \label{eq:hodge}
    h^{1,1}=256\,,\quad h^{2,1}=0\,,\quad h^{3,1}=289384\,,\nonumber\\
    h^{2,2}=44+4h^{1,1}-2h^{2,1}+4h^{3,1}=1158604\,,\nonumber\\
    \chi=6(8+h^{1,1}-h^{2,1}+h^{3,1})=1737888\,.
\end{gather}
Notice that there are many more fourfolds with the
same $\chi$, but they all have much larger
$h^{1,1}$ and are harder to analyze, while they
very probably do not give larger $q_\mathrm{max}$, as discussed in
Section~\ref{subsec:bound}.

$\hat Y$ is a \cy hypersurface in an ambient
toric fivefold, a (singular) weighted projective
space $\mathbb P^{1,80,492,1148,1722}$ \cite{Scholler:2018apc}. It can also
be understood as a generic elliptic fibration over
a toric base $B$ to be specified below. The
equivalence of the two descriptions is shown in
Appendix \ref{sec:equivalence}. Now, $B$ can be
described as a $B_2$-bundle over $\mathbb P^1$,
where $B_2$ is a toric surface characterized by a
closed cycle of divisors (or rays in the 2D toric
fan) with self-intersection numbers
$0,6,-12//-11//-12//-12//-12//-12//-12//-12//-12$,
where $//$ represents the chain
$-1,-2,-2,-3,-1,-5,-1,-3,-2,-2,-1$ \cite{MorrisonTaylorToric,TaylorWangVacua}. Its toric rays
$v_\alpha\in \mathbb Z^2$ can be taken to be\footnote{Note that in
  \cite{TaylorWangVacua}, the indices on $v_\alpha, w_\alpha$ etc. are
taken to be  roman indices $i$; here to avoid confusion we use the
appropriate base divisor index notation $\alpha$, although when there
is possible ambiguity with integer indices $i$ indexing Cartan
divisors
as in $D_i$, we put the index as a superscript or use alternative
explicit non-integer notation.}
\begin{equation}
    v_{1}=\left(-1,-12\right)\,,\quad v_{2}=\left(0,1\right)\,,\quad...\,,\quad v_{99}=\left(0,-1\right)\,,
\end{equation}
where the intermediate rays are determined by
$v_{\alpha-1}+v_{\alpha+1}+C_{\alpha}^{2}v_{\alpha}=0$ and $C_{\alpha}^{2}$
is the self-intersection number of the divisor
corresponding to $v_\alpha$, starting at
$C_1^2=0,C_2^2=6$. Then the 3D toric fan of $B$ is
given by the rays $w_\alpha$:
\begin{equation}
    w_{0}=\left(0,0,1\right)\,,\quad w_{1\leq \alpha\leq99}=\left(v_{\alpha},0\right)\,,\quad w_{100}=\left(80,468,-1\right)\,,
\end{equation}
where $(80,468)=4v_{19}$ is the twist of the
$B_2$-bundle. We denote the corresponding divisor
classes to be $D^\alpha$, where the superscript integer
indexing the base divisors is
distinguished from the subscript for exceptional
divisors as mentioned above; when we use $\alpha$ as a subscript where
Cartan indices $i$ are also possible, as in,
e.g., $S_{i \alpha}$ we use non-integer notation for the $\alpha$'s. The cones of the fan are given by
$\left(w_{0},w_{\alpha},w_{\alpha+1}\right),\left(w_{100},w_{\alpha},w_{\alpha+1}\right)$ for $1\leq \alpha\leq98$,
as well as $\left(w_{0},w_{99},w_{1}\right),\left(w_{100},w_{99},w_{1}\right)$.

The local geometry on divisor $D^0=D^{100}$ is
clearly $B_2$, while that on divisors $D^{1\leq \alpha\leq 99}$ are all Hirzebruch surfaces
$\mathbb F_n$. In particular, we have
$h^{1,1}(D^{1\leq \alpha\leq 99})=2$. The intersection numbers on
$D^{1\leq \alpha\leq 99}$ are then determined by
$n$ only. Note that since the twist is along $v_{19}$,
the local geometry on $D^{19}$ is $\mathbb F_0$,
which has the smallest intersection numbers among all $\mathbb F_n$.

Some of
the divisors $D^{1\leq \alpha\leq 99}$ have sufficiently
negative normal bundles in $B$ that the
elliptic fibration is forced to be singular to certain
degrees, and nonabelian gauge factors
automatically arise on these divisors. Such  \emph{rigid} or \emph{geometrically
non-Higgsable} gauge groups are
present throughout the whole set of moduli space
branches associated with  elliptic \cy's over such a base \cite{MorrisonTaylorClusters,MorrisonTaylor4DClusters}. As
a result, these gauge groups cannot be broken by
any geometric deformation (corresponding to
Higgsing from the low-energy perspective), while
they can still be broken by fluxes. The method for
determining the rigid gauge groups in 4D F-theory
models has been described in \cite{MorrisonTaylor4DClusters}, and here we
summarize the result applied in this type of case where the base $B$
is a $B_2$ bundle over $\P^1$. The divisor $D^{1\leq \alpha\leq 99}$ supports $E_8$ if $C_{\alpha}^{2}=-11,-12$, $F_{4}$
if $C_{\alpha}^{2}=-5$, $G_{2}$ if $C_{\alpha}^{2}=-3$, and
$\SU\left(2\right)$ if $C_{\alpha}^{2}=-2$ and
intersects with a $G_{2}$ gauge divisor. Therefore,
the full gauge group is
\begin{equation} \label{eq:totalgaugegroup}
    E_{8}^{9}\times F_{4}^{8}\times\left(G_{2}\times\mathrm{SU}\left(2\right)\right)^{16}\,.
\end{equation}
In particular, there is a $G_2$ factor supported on $D^{19}$.

Note that there may be codimension-2 $(4,6)$ singularities
localized on divisors supporting $E_8$ factors. By computing
the normal bundles on divisors, one can check that there are
four irreducible components of codimension-2 $(4,6)$ loci on
$D^3$ (with $C_3^2=-12$) and one on $D^{15}$ (with $C_{15}^2=-11$).
To remove these singularities, non-toric blowups must be performed,
contributing $5$ to the $h^{1,1}$ in Eq.\ (\ref{eq:hodge}).
These singularities are associated with extra strongly coupled
(probably conformal)
sectors that have not been well understood \cite{HeckmanMorrisonVafa,DelZotto:2014hpa,Apruzzi:2018oge}.
Nevertheless, these sectors are decoupled from the gauge sectors
we are studying and should not affect our analysis.

\subsection{Flux background}
\label{subsec:fluxbackground}
Now we would like to break some gauge factors in
Eq.\ (\ref{eq:totalgaugegroup}) to get an exotic
$\U(1)$ using vertical flux breaking. Since
vertical flux breaking must decrease the rank of
the gauge group, we cannot have breaking like
$\SU(2)\rightarrow\U(1)$. By Eq.\ (\ref{eq:rs}) with
$r=2$ for all gauge factors, we then see that the
only available breaking is $G_2\rightarrow\U(1)$.
One may naively consider a breaking like
$G_2\times\SU(2)\rightarrow\U(1)$ where the $\U(1)$
is a combination of Cartan generators from both
gauge factors, since a $G_2$ gauge divisor always
intersects with an $\SU(2)$ gauge divisor. It can
be shown that, however, such breaking violates an
analogous version of Eq.\ (\ref{eq:rs}).

Here we reach one of the main points in this section:
we can minimize the intersection numbers on the
gauge divisor by performing the flux breaking on
$D^{19}$, which is locally $\mathbb F_0$ and
supports a $G_2$. This crucial feature is why we
study the fifth largest $h^{3,1}$ and $\chi$ in the
KS database but not one of the \cy's with even larger $\chi$.

Let us specify more details on $D^{19}$. The only
$D^i$'s that intersect with $D^{19}$ are
$D^0=D^{100},D^{18},D^{20}$. The curves on $D^{19}$
are then spanned by $D^0\cdot D^{19}=D^{100}\cdot D^{19}$ and $D^{18}\cdot D^{19}=D^{20}\cdot D^{19}$. Following the notation
in Section \ref{subsec:simpleexample}, we denote
$\Sigma=D^{19}, S=D^0, F=D^{18}$. The only
nontrivial intersection number on $\Sigma$ is
$\Sigma\cdot S\cdot F=1$. Now to break $G_2\rightarrow\U(1)$, we turn on nonzero $\phi_{i\alpha}$ such that
\begin{equation}
    a\Theta_{1\alpha}+b\Theta_{2\alpha}=0\,,
\end{equation}
for all $\alpha=S,F$, where the index $i$ is the
Cartan index for the $G_2$, and integers $a,b$ are
coprime. The labels $1,2$ correspond to
\begin{equation}
    \kappa^{ij}=\left(
    \begin{array}{cc}
        6 & -3\\
        -3 & 2
    \end{array}\right)\,.
\end{equation} 
If $(a,b)$ is not along any root of $G_2$,
all the roots of $G_2$ are broken and the remaining
gauge group is $\U(1)$ with generator
$T=aT_{1}+bT_{2}$. The flux constraint is solved by
\begin{equation}
    \left(\phi_{1S},\phi_{2S},\phi_{1F},\phi_{2F}\right)=\left(\left(a-2b/3\right)n_{S},\left(2a-b\right)n_{S},\left(a-2b/3\right)n_{F},\left(2b-a\right)n_{F}\right)\,,
\end{equation}
where $n_S,n_F$ are flux parameters to be chosen.
To satisfy flux quantization, we see that $n_S,n_F$
must be multiples of 3 unless $b$ is a multiple of 3.
When $b$ is a multiple of 3, one can show that
$(a-2b/3)$ and $(2a-b)$ must be coprime and
$n_S,n_F$ must be integer. Since the size of $\phi$
has been bounded, to ensure the most exotic choice
of $(a,b)$ we should assume $b$ as a multiple of 3
and integer $n_S,n_F$. Now we turn to primitivity;
as in Section \ref{subsec:simpleexample},
only the K\"ahler
form of $D^{19}$ contributes in Eq.\ 
(\ref{eq:intprimitivity}) with $D_{i}$ being the
exceptional divisors from the $G_2$. Therefore, we require
$n_S,n_F$ to be both
nonzero and have opposite signs.

With the above information, we can easily calculate the tadpole from this flux:
\begin{equation}
    \frac{1}{2}\int_{\hat Y} G_{4}\wedge G_{4}=-2\left(a^{2}-ab+\frac{b^{2}}{3}\right)n_{S}n_{F}\,.
\end{equation}
We see that to minimize the tadpole and satisfy
primitivity, we should choose e.g.
$(n_S,n_F)=(1,-1)$. This ensures the tadpole
to be positive. Then Eq.\ (\ref{eq:tadpole}) becomes
\begin{equation} \label{eq:g2tou1tadpole}
    a^{2}-ab+\frac{b^{2}}{3}\leq36206\,.
\end{equation}
To maximize the $\U(1)$ charges, we should choose
$(a,b)$ such that the above is the closest to
saturation. The ratio between $a$ and $b$ is now
determined by the matter spectrum charged under the
$G_2$. Let us first focus on the adjoint
$\mathbf{14}$ of $G_2$. After the breaking, the
W-bosons become charged singlets with the $\U(1)$
charges
\begin{equation}
    a,\,b,\,3a-b,\,2a-b,\,a-b,\,3a-2b\,,
\label{eq:general-g2-spectrum}
\end{equation}
and their conjugates. There are also two uncharged
singlets. To find out the largest possible
$q_\mathrm{max}$, we then maximize one of the
charges in the above, subject to the tadpole constraint
and the assumption of $b$ being multiple of 3. It
turns out that there are multiple choices giving
the same largest $q_\mathrm{max}$. For example,
$(a,b)=(329,657)$ (or $(\phi_{1S},\phi_{2S})=(-109,1)$)
gives the largest $b=657$,
with the full set of $\U(1)$ charges from the adjoint being
\begin{equation}
    q=1,\, 327,\, 328,\, 329,\, 330,\, 657\,.
\end{equation}
Therefore, we have reached one of
the main results of this paper, a $\U(1)$ 4D F-theory model with
\begin{equation}
    q_{\max}=657\,.
\end{equation}

To complete the discussion, we still need to look
at other representations. There is also
bifundamental matter $(\mathbf{7},\mathbf{2})$
charged under $G_2\times\SU(2)$ before breaking \cite{MorrisonTaylorClusters}. It
breaks into representations of $\SU(2)\times\U(1)$
after the breaking, so the $\U(1)$ is still coupled
to other gauge factors. One can, however, turn on
one more unit of vertical flux to break the
adjacent $\SU(2)$ completely. Then the
bifundamental also breaks into $\U(1)$ charged
singlets and the $\U(1)$ we constructed is fully
decoupled. The same calculation as above shows that
the bifundamental only gives a subset of $\U(1)$
charges coming from the adjoint, with the maximum being $q=329$.

It is informative to study the chiral spectrum of these large
charges. Interestingly, Eq.\ (\ref{eq:chi}) implies that the chiral
indices from the adjoint are proportional to $(n_S+n_F)$, hence
vanish in the above example. In particular, it means that the
charge $q=657$ must belong to vector-like matter, which is not exactly
massless if including interactions in the superpotential. A careful
calculation using the approach of \cite{BeasleyHeckmanVafaI}
shows that the multiplicity of vector-like $\mathbf 1_{657}$ in this model is indeed nonzero.
On the
other hand, there is chiral matter from the bifundamental. Since the
adjacent $\SU(2)$ is completely broken, we can effectively consider
two copies of $\mathbf 7$ localized on $C_{\mathbf 7}=D^{18}\cdot D^{19}=\Sigma\cdot F$. Eq.\ (\ref{eq:chi}) then gives the following chiral spectrum:
\begin{equation}
    438\times\mathbf 1_1+220\times\mathbf 1_{-328}+218\times\mathbf 1_{329}\,,
\end{equation}
which is again anomaly-free as expected. Therefore, if we restrict
to the truly massless chiral fields only, $q_\mathrm{max}$ is not as
large as 657. There are still ways to go beyond $q=329$ for chiral
fields. For example, in the same model as above, we can choose
$(n_S,n_F)=(2,-1)$ instead. Then there are chiral fields from the
adjoint, and the same calculation of $q_\mathrm{max}$ from the adjoint
gives $q_\mathrm{max}=465\simeq 657/\sqrt{2}$.

One may naively expect, from the low-energy perspective,
that we can
give the above massless chiral fields a vacuum expectation value to Higgs the symmetry to
a discrete abelian group
$\U(1)\rightarrow\mathbb Z_k$.
The above example then suggests that $k$  could be as large as 465 for
such discrete symmetries in 4D.\footnote{We thank
Paul Oehlmann for raising this point.}
This is much larger than
the largest size  $\mathbb Z_6$ currently known 
(\cite{Anderson:2019kmx} and references therein)
for discrete gauge
symmetries from 
Tate-Shafarevich/Weil-Ch\^atelet groups of smooth elliptic
\cy threefolds or fourfolds \cite{Braun:2014oya,Morrison:2014era}.
Nevertheless in 4D, there are various Yukawa couplings involving these
chiral fields, which can induce a potential and stabilize these
vacuum expectation values. As a result, although here we do not demonstrate
it explicitly, we expect that such Higgsing to discrete gauge symmetries is
not possible in our setup.

One should be reminded that this kind of $\U(1)$ model
is very rare in the 4D F-theory landscape, as we
have almost saturated the tadpole bound, and
arranged all fluxes to be along several specific
directions. In particular, with these constructions there is almost no room
to turn on horizontal flux for moduli
stabilization.
A generic $\U(1)$ model is expected to have
fluxes spreading over many directions, with only
a small amount of flux along each direction, hence giving small $\U(1)$ charges.

\subsection{Towards an upper bound}
\label{subsec:bound}
One important question regarding $\U(1)$ charges of
massless fields is
whether an upper bound on such $q$ exists, and if so what that upper
bound is. With our
current technologies, it seems impossible to precisely determine
the value of the upper
bound with certainty, since the lists of elliptic \cy fourfolds
$\hat Y$ and bases $B$, as well as tools for
building $\U(1)$ models, are rather limited.  One can
certainly attempt to seek models that
exceed our result $q_\mathrm{max}=657$. Here, however, we
provide some heuristic reasons for why we expect our
result may give, or at least be close to, an actual upper bound
on $q$ within the 4D
F-theory landscape.
\begin{itemize}
    \item The most straightforward way to find
    other large $\U(1)$ charges is to generalize
    the method of vertical flux breaking to other known
    geometries. There are four known \cy
    fourfolds with
    $h^{3,1}$ and $\chi$ larger than those in our
    model  (these are all in the KS database; Euler characters of, e.g.,
     CICY fourfolds are much smaller, with $\chi \leq 2610$
\cite{Gray:2014fla}). The bases from these fourfolds are also $B_2$-bundles
    over $\mathbb P^1$, with the same $B_2$
but with different twists \cite{TaylorWangVacua}.
    Note that none of these twists are along a
    $G_2$ gauge divisor, so the local geometries of
    the $G_2$ gauge divisors are never as simple as
    $\mathbb F_0$. In fact, the same construction
    as in our model needs to be done on $\mathbb F_3, \mathbb F_6, \mathbb F_9, \mathbb F_{12}$
    respectively when $h^{3,1}$ and $\chi$
    increase. Therefore, the increase of
    intersection numbers on $G_2$ gauge divisors
    surpasses the slight increase of $\chi$, and
    leads to lower $q_\mathrm{max}$. The geometries
    with the same $\chi$ but lower $h^{3,1}$ in the
    KS database have much larger $h^{1,1}$ and are
    harder to analyze. Although we do not have any
    quantitative statements, the general
    expectation is that these geometries contain many more rigid
    gauge groups, and the divisor geometries are generically
    more complicated with higher $h^{1,1}$.  Due to
    such complexity, we may not
    expect there to be a $G_2$ gauge divisor as simple as
    $\mathbb F_0$. Even if there is such a gauge divisor, by
    the same construction the resulting charge should not be
    significantly larger than 657.
    
    \item In principle, there may be elliptic \cy fourfolds with much larger $\chi$ than that in
    our model, thus potentially giving much larger
    $q_\mathrm{max}$. 
From what is known of the structure of elliptic threefolds and
fourfolds, however, it seems unlikely that $\chi$ of any elliptic
fourfold
can exceed those that are known and mentioned above.  While this
cannot be proven rigorously, we summarize some arguments for this here.
    The situation for elliptic threefolds is fairly clear: there are
a finite number of
    elliptic \cy threefolds \cite{GrossFinite}
and all of the allowed bases have
    been classified by the minimal model program \cite{Grassi1991}.
The elliptic \cy threefold with the largest $h^{2, 1}$ is known to
be the generic
elliptic fibration over $\mathbb F_{12}$ \cite{Taylor:2012dr}, which has the largest known
(absolute value of) Euler character $|\chi| = 960$.  The distinctive ``shield'' shape of the Hodge
numbers for all toric hypersurface \cy threefolds has 3 peaks with maximum
$h^{1, 1}+h^{2, 1}$, which are all
realized by elliptic fibrations over toric bases.
(Because of the alternating signs in the Euler character,  $h^{1, 1}
+h^{2, 1}$ may be a better proxy for the Euler character of fourfolds
than the threefold Euler character $2 (h^{1, 1}-h^{2, 1})$.)
A systematic classification of the allowed bases, including all toric
bases
\cite{MorrisonTaylorToric} and non-toric bases  giving \cy threefolds with
$h^{2, 1}\geq 150$ \cite{TaylorWangNon-toric} shows that the toric hypersurface \cy threefolds
in the KS database \cite{Kreuzer:2000xy} accurately
capture the boundary of the set of possible Hodge numbers. 
In particular, there is known to be
no \cy threefold with larger (absolute value
of) Euler character or $h^{1, 1}+h^{2,1}$  among generic elliptic
fibrations with  $h^{2, 1}\geq 150$ over any base surface.
This gives extremely strong (but not airtight) evidence that the
largest values of the Euler character and $h^{1, 1}+h^{2,1}$ for
elliptic \cy threefolds are realized by elliptic fibrations
over toric bases and are found at the boundary points of the KS database.

While    it is far from clear whether the analogous
    statement is true for fourfolds, it seems very plausible that this
    should be true. 
The
 shape of the Hodge shield (in $h^{1, 1},h^{3,1}$) for CY fourfolds
 takes a very similar, although more spiky, form to that for
 threefolds, with again 3 prominent cases with maximum $h^{1, 1}+h^{3,1}$
\cite{Kreuzer:1997zg,Scholler:2018apc}, corresponding again to the
largest known CY fourfold Euler characters.
From the
    perspective of the analogous minimal model program (the Mori program)
    for threefold bases, it is expected that the largest $h^{3,1}$
    will come from a minimal threefold base that is either Fano, a 
$\mathbb P^1$ bundle over a surface $B_2$, or a
    $B_2$-bundle over $\mathbb P^1$.  The last of these classes seem to
    give the largest possible values for $h^{3,1}$ and $\chi$
    \cite{Klemm_1998,Halverson:2015jua,TaylorWangVacua}, and  as we
    have discussed here the bases we have used with large $h^{3,1}$
    are all $B_2$ bundles over $\mathbb P^1$.  If the fourfold case
    follows the better understood pattern of geometries for
    threefolds, these are indeed the elliptic CY fourfolds with
    largest $\chi$.
As for threefolds,
currently most known elliptic
    \cy fourfolds come from hypersurfaces in toric
    ambient spaces or elliptic fibrations on toric
    threefold bases \cite{TaylorWangMC,HalversonLongSungAlg,TaylorWangLandscape}, so elliptic \cy fourfolds with
    larger $\chi$, if they exist, would very probably
    involve non-toric constructions. 
It has recently been
    noticed that, unlike in 6D, non-toric
    constructions of elliptic \cy fourfolds and threefold
    bases seem to give an additional large class of
    4D F-theory models \cite{Braun:2014pva,LiFluxbreaking} with
    qualitatively novel features.
The extent
    of such geometries is certainly an open
    question, although, as for elliptic \cy threefolds, it is known that
    the number of topological types of
elliptic \cy fourfolds is finite up to
birational equivalence \cite{di2021birational}.
From analogy with CICY fourfolds, however, where the Euler characters
as mentioned above are generally much smaller than those of toric
hypersurfaces, and from experience with non-toric bases for elliptic
threefolds
\cite{TaylorWangNon-toric}, it seems natural to expect that  non-toric
bases will not give larger Hodge numbers or Euler characters than the
examples already known.  Thus,
we think that it is not unreasonable to
    believe that there may be no fourfolds
    with $\chi$ significantly larger than that in our model.
Rigorous results in these directions, however, are clearly an important
direction for further work.

    \item
It is natural to consider $\U(1)$ factors from
    breaking of gauge groups other than $G_2$, but
such $\U(1)$'s are unlikely to give larger
    $q_\mathrm{max}$. First, consider $\U(1)$ factors arising
from
    gauge groups with higher rank. Eq.\     (\ref{eq:rs}) then requires $h^{1,1}(\Sigma)>2$, so we cannot use bases
    as simple as a $B_2$-bundle over $\mathbb P^1$ in the same
    way,
    and we are forced to consider  more
    complicated divisor geometries, which may lead to tighter tadpole
    constraints, as discussed previously.  Moreover, from the
    calculation in our model, it seems that the
    optimization of $q_\mathrm{max}$ can be done by
    localizing almost all the
    flux $\phi_{i\alpha}$ on one of the exceptional
    divisors.
    Therefore, the presence of additional Cartan
    directions should not significantly change the
    optimization process. The third reason arises when considering gauge groups with any rank: $G_2$ has the
    most exotic root vectors due to the presence of
    $3$ in the components, so generically the
    resulting $\U(1)$ charges from $G_2$ are larger
    than those from other gauge groups. All these
    reasons lead us to expect that the $G_2$ breaking is
    likely to give the largest $q_\mathrm{max}$.
    
    \item Finally, the possibilities of $\U(1)$
    models provided by vertical flux breaking
    clearly greatly exceed other available methods in
    literature. In contrast to the construction analyzed
    here,
    realizing $\U(1)$ with additional rational
    sections and nontrivial Mordell-Weil group
    relies heavily on the global geometry. Given the
    difficulty of building such models even with charges up to $q =
    6$, as
    summarized in Section \ref{sec:Intro}, it is
    reasonable to expect that such a construction
    can never exceed, or even approach, the result found here. Moreover, as
    discussed in Section \ref{subsec:breaking},
    vertical flux breaking provides even more
    flexibility than Higgsing in field theory,
    which also corresponds to geometric
    deformation in F-theory. Therefore, we expect
    our result to exceed any charges obtained from
    Higgsing arguments.
\end{itemize}

Readers should be warned that all the above are
only heuristic arguments. It is possible that
some of these arguments are not true in general,
and that larger, or even much larger, $q_\mathrm{max}$ can be found in
F-theory. Although
F-theory is so far the most promising approach to
exploring global aspects of the nonperturbative string landscape, we also cannot exclude
the possibility that there are compactifications
in other corners of string theory that give rise
to even larger $q_\mathrm{max}$. We have not
explored non-geometric or non-supersymmetric
constructions in this regard at all. Clearly much more work needs
to be done to rigorously construct an upper bound on for $q$, but
hopefully the work and arguments presented here provide a starting
point for further analysis.

\section{Coupling to other gauge groups}
\label{sec:coupling}

So far we have  focused on a single $\U(1)$
gauge factor, decoupled from other gauge
groups; it is also interesting to study a $\U(1)$ factor
coupled to other gauge groups. In particular, this
is phenomenologically interesting since the
hypercharge $\U(1)$ and various $\U(1)$ extensions
to the Standard Model have this structure. 
Although the hypercharge $\U(1)$ in the
Standard Model cannot be obtained from purely vertical
flux breaking \cite{LiFluxbreaking}, one may still apply vertical
flux breaking to build various $\U(1)$ extensions. On
the theoretical side, we expect that
$q_\mathrm{max}$ may exceed the value of 657 found above when the
$\U(1)$ is coupled to other gauge groups, since there are many more
possible breaking scenarios with other gauge factors and more parameters.
A full analysis of this problem is
beyond the scope of this paper.  Here we
simply demonstrate  a single example with a slightly larger $q_\mathrm{max}$ than that in
Section \ref{subsec:fluxbackground}.
While there may be larger values possible, for similar reasons to
those discussed in Section \ref{subsec:bound}, we do not expect
enormously larger values for $\U(1)$ charges even when other gauge
factors are included.
Note that in many cases when the $\U(1)$ factor couples to one or more
nonabelian factors, the global structure of the group may have a quotient
by a discrete component of the center, such as in the Standard Model
group where the global structure seems likely to be $(SU(3) \times
SU(2) \times U(1))/\Z_6$ (see, e.g., \cite{LawrieEtAlRational,GrimmKapferKleversArithmetic,CveticLinU1,TaylorTurnerGeneric,TaylorTurner321}).  In such cases it is often conventional to
use fractional values for $\U(1)$ charges, as is often used in the
(unbroken) Standard Model.  
In our discussion here, as mentioned in
a footnote in Section \ref{sec:Intro}, we always treat
$\U(1)$ charges as integers, with
the minimal $\U(1)$ charge being $q = 1$.  With this normalization,
while the approach of flux breaking provides $\U(1)$ factors with much
larger charges than those available  directly from F-theory Weierstrass
models (as analyzed in, e.g., \cite{LawrieEtAlRational}), we
expect bounds of similar magnitude on $q_\mathrm{max}$ in the presence of
nonabelian factors to those found above for pure $\U(1)$ factors.

Following the construction in Section
\ref{sec:u1}, here we build a similar $\U(1)$
model with $q_\mathrm{max}=672>657$, but where the
$\U(1)$ is coupled to an $E_6$. This time, we use
the known \cy fourfold with the largest $h^{3,1}$
and $\chi$. It has been argued that this geometry
plausibly supports the most flux vacua in the 4D
F-theory landscape \cite{TaylorWangVacua}. The geometry is the same as
that in Section \ref{subsec:geometry} except the
twist of the $B_2$-bundle. To be precise, we
replace the toric ray $w_{100}$ of $B$ in Section
\ref{subsec:geometry} by
\begin{equation}
    w_{100}=(84,492,-1)\,,
\end{equation}
where $(84,492)=12v_{15}$ is the twist. The
fourfold has $\chi=1820448$. The rigid
gauge groups are the same as before. We notice
that the divisor $D^{15}$ now has local geometry
$\mathbb F_0$ and supports a rigid $E_8$. Although
we cannot construct a single $\U(1)$ gauge group,
Eq.\ (\ref{eq:rs}) allows us to perform the
breaking $E_8\rightarrow E_6\times \U(1)$ on this
divisor.\footnote{As noted above, there are
codimension-2 $(4,6)$ singularities on $D^{15}$ associated
with extra strongly coupled sectors. Unless
these sectors have direct conflict with vertical
flux breaking (which we do not see immediately),
they should be irrelevant to the matter coming
from the adjoint of $E_8$, which is localized on
the bulk of $D^{15}$ instead of on matter curves.}

\begin{figure}[t]
\centering
\includegraphics[width=0.5\columnwidth]{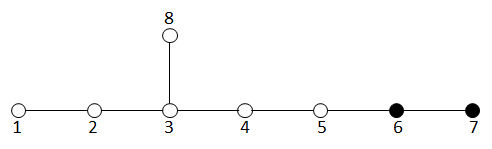}
\caption{The Dynkin diagram of $E_8$. The Dynkin node labelled
$i$ corresponds to the exceptional divisor $D_i$. The solid
nodes are the ones we break by vertical flux,
while we preserve a linear combination of their
corresponding Cartan generators. The unbroken
nodes form the Dynkin diagram of $E_6$.}
\label{dynkine8}
\end{figure}

Now the calculation is similar to that in Section
\ref{subsec:fluxbackground}. The breaking can be done by imposing (see Figure \ref{dynkine8})

\begin{equation}
    \Theta_{1\alpha}=\Theta_{2\alpha}=\Theta_{3\alpha}=\Theta_{4\alpha}=\Theta_{5\alpha}=\Theta_{8\alpha}=a\Theta_{6\alpha}+b\Theta_{7\alpha}=0\,,
\end{equation}
where $\alpha$ stands for $S=D^0$ and $F=D^{14}$,
and $a,b$ are coprime integers.
We further require that there is no $E_8$ root
with the sixth and seventh components along
$(a,b)$. Now the above flux constraints are solved by
\begin{gather}
    (\phi_{1S},\phi_{2S},\phi_{3S},\phi_{4S},\phi_{5S},\phi_{6S},\phi_{8S})=(2,4,6,5,4,3,3)(a-2b)n_S\,,\nonumber\\
    \phi_{7S}=(2a-3b)n_S\,,
\end{gather}
and similarly for $\phi_{iF}$. Since $(a-2b)$ and
$(2a-3b)$ are also coprime, $n_S,n_F$ must be
integers. Primitivity still requires $n_S,n_F$ to
have opposite signs, and we choose $(n_S,n_F)=(\pm 1,\mp 1)$ to minimize the tadpole as before. Then
Eq.\ (\ref{eq:tadpole}) becomes
\begin{equation} \label{eq:e8tofe6tadpole}
    a^2-3ab+3b^2\leq37926\,.
\end{equation}
Note the similarity to Eq.\ (\ref{eq:g2tou1tadpole}).

We now look at the matter spectrum. The adjoint
$\mathbf{248}$ of $E_8$ breaks into $E_6$
fundamentals $\mathbf{27}$ (and their conjugate)
and singlets that are charged under the $\U(1)$.
To be precise, the charged representations are
\begin{equation}
    \mathbf{27}_{a},\,\mathbf{27}_{a-3b},\,\mathbf{27}_{-2a+3b},\,\mathbf{1}_{3b},\,\mathbf{1}_{3a-3b},\,\mathbf{1}_{3a-6b}\,,
\end{equation}
and their conjugates. 
Similar to the example in Section \ref{subsec:fluxbackground},
the flux induces no chiral spectrum and the above representations
belong to vector-like matter only.
The maximum possible charge can be
obtained by maximizing the charge of one of the
singlets while satisfying Eq.\ 
(\ref{eq:e8tofe6tadpole}). For example, $(a,b)=(2,-111)$ gives
maximum $3a-6b=672$ and the charged representations are
\begin{equation}
    \mathbf{27}_{2},\,\mathbf{27}_{335},\,\mathbf{27}_{-337},\,\mathbf{1}_{-333},\,\mathbf{1}_{339},\,\mathbf{1}_{672}\,,
\end{equation}
and their conjugates. Therefore, we have reached
the result $q_\mathrm{max}=672$ for $\U(1)$
coupled to $E_6$. Note that, as discussed above, the charge is 672 by
the definition we are using here where all $\U(1)$ charges are integers, while
the spectrum is invariant under the $\mathbb Z_3$
center of the gauge group. Therefore, it is also
natural to normalize the charges with units of
$1/3$, in which units the maximum charge would become
$q_\mathrm{max}=224$. 
One should thus be careful when
interpreting the result in this Section, while the
same ambiguity does not occur in Section \ref{sec:u1}.

It is worth emphasizing again that these models
with large $\U(1)$ charges require very
non-generic flux configurations, and small $\U(1)$
charges are exponentially preferred in the
landscape (assuming a measure where all flux configurations satisfying
the tadpole constraint are equally weighted). Heuristically, if one collects all the
$\U(1)$ charges of massless states across  the
landscape, one may expect a distribution that
peaks near $q=0$ and decays as $q$ grows \cite{Andriolo:2019gcb}. This
matches the expectation from phenomenology where
the $\U(1)$ charges are always small.

We see that no matter how large the
charges are, some Yukawa couplings such as
$\mathbf{27}^3$ in the above models are allowed to  be possible
from the structure of the unbroken gauge group,
while some other couplings are indeed forbidden by
the inclusion of exotic $\U(1)$ charges. The above
breaking can be straightforwardly generalized to
conventional grand unified theories such as
$E_6\rightarrow \SU(5)\times\U(1)$. Therefore,
vertical flux breaking can be useful in
phenomenological model building, in which the
$\U(1)$ extension may naturally explain issues
like proton decay \cite{Marsano:2009wr,Grimm:2010ez}. In particular, such
constructions can be easily incorporated into the
recently proposed natural construction of Standard
Model-like structure in F-theory \cite{Li:2021eyn,LiFluxbreaking}, as vertical flux
breaking is used in both contexts.

\section{Conclusion}
\label{sec:Con}

In this paper, we have studied $\U(1)$ charged
massless fields in string theory. 
While the completeness hypothesis
\cite{Polchinski:1998rr,Banks:2010zn,Harlow:2018tng} suggests that
states should exist with all possible charges under a U(1) gauge
field, in general massless or light fields have small charges.
Even in 6D, 
the upper bound
on  possible $\U(1)$ charges $q$ for  massless fields
is not completely understood, with very few
explicit string theory constructions of $\U(1)$
models available, and in 4D the question is addressed very little in
the literature. Using the formalism of vertical
flux breaking, we have efficiently constructed 4D F-theory models containing
$\U(1)$ charges for light fields that can be
as large as $q=657$, and massless chiral fields with charges that can
be at least as large as $q = 465$.
The string theory construction is
fully explicit, using fourfolds with large Euler
characteristics in the KS database and exotic flux
background. While a rigorous proof is far from
complete, we have some reasons to believe that our
result gives a plausible upper bound for $\U(1)$
charges in the 4D F-theory landscape, when the
$\U(1)$ is decoupled. We have found that the
$\U(1)$ charges can become slightly larger when the
$\U(1)$ factor is coupled to other gauge groups,  although we expect a similar
upper bound when for properly normalized charges.

It is worth emphasizing that the St\"uckelberg mechanism is also
available in, for example, type IIB string theory (see e.g.
\cite{Cvetic:2012xn} and the references therein), and one can construct
similar $\U(1)$ models in such a perturbative setup. The nonperturbative
nature of F-theory, however, allows us to explore a much wider range of
compactifications, in particular those containing exceptional gauge
groups as in our examples. As a result, F-theory opens up new corners in
the string landscape by raising the $\U(1)$ charges that can be explicitly constructed
to a much higher value. It remains interesting to compare the upper
limits from both F-theory and perturbative string theories.

These results lead in a number of interesting directions. First, as
described in Section \ref{sec:Intro}, this class of constructions
may give new insights in the context of the
Swampland program, as  it greatly expands the
view on which $\U(1)$ charged massless fields in
the low-energy theory can be consistently coupled
to gravity. 
It would be interesting to understand better how these 4D
constructions of theories with light fields having large U(1) charges
fit with the related 6D analyses of
automatic enhancement and of massless charge sufficiency and
the completeness hypothesis
\cite{Raghuram:2020vxm,Cvetic:2021vsw,Morrison:2021wuv}.
More generally, we have found that the
formalism of vertical flux breaking 
leads
to large classes of new F-theory models ranging from exotic
$\U(1)$ charges to natural Standard Model-like
constructions. 
At the same time, the large charges we have found here are expected to
be exponentially rare in any natural measure on the landscape, and
this work motivates a more careful study of the impact of the
distribution of fluxes on the structure of the gauge group and matter
content. While this approach using flux breaking is certainly not completely new
in the literature, it has not been fully utilized until
now. 
We hope that this formalism will lead to many more
exciting results in  F-theory  constructions of 4D supergravity
models.
 Finally, this perspective offers new
ways of thinking about $\U(1)$ extensions in the
Standard Model or grand unified theories. As shown
in Section \ref{sec:coupling}, the $\U(1)$ charges
we get, albeit exotic, are still potentially
relevant to particle phenomenology. 
It is interesting that the possible
appearance of such charges is supported from the UV perspective by explicit
string theory constructions.

We hope that future work will lead to a solidification of the
arguments for the upper bound on $q$, and the application of
this construction of exotic $\U(1)$ charges to
broader setups as mentioned in the previous
paragraphs.

\acknowledgments{We would like to thank James Gray, Patrick Jefferson,
  Manki Kim,
  Paul Oehlmann, and Andrew Turner for helpful discussions. This work
  was supported by the DOE under contract \#DE-SC00012567.}

\appendix

\section{Various descriptions of the geometry}
\label{sec:equivalence}

In this Appendix, we compare two descriptions of
the geometry in Section \ref{subsec:geometry}, and
show that they are equivalent. One description is
a generic elliptic fibration over a given
threefold base, and another one is the
anticanonical hypersurface in a 5D (singular)
weighted projective space.

Let us provide more details on the elliptic fibration. Since the elliptic curve is the \cy hypersurface in weighted projective space $\mathbb P^{2,3,1}$, the elliptic fibration on a toric base can also be written as the \cy hypersurface in a 5D toric ambient space. The toric ambient space is a $\mathbb P^{2,3,1}$ fibration over the base, given by the following toric rays:
\begin{equation}
    (w_i,-2,-3)\,,\quad(0,0,0,1,0)\,,\quad(0,0,0,0,1)\,.
\end{equation}
The resulting fivefold is singular due to the
presence of rigid gauge groups. Those
singularities on gauge divisors can be resolved by
adding ``tops'' into the toric fan \cite{Candelas:1996su}. We do not
describe the details here, but one can show that
after adding all the tops from Eq.\ 
(\ref{eq:totalgaugegroup}), the convex hull of the
toric fan is a reflexive polytope with vertices
\begin{gather}
    \left(0,0,1,-2,-3\right)\,,\quad\left(-1,-12,0,-2,-3\right)\,,\quad\left(0,1,0,-2,-3\right)\,,\nonumber\\\left(0,0,0,1,0\right)\,,\quad\left(0,0,0,0,1\right)\,,\quad\left(80,468,-1,-2,-3\right)\,,\quad\left(42,246,0,-2,-3\right)\,.
\end{gather}
Notice again that $(80,468)=4v_{19}$ and $(42,246)=6v_{15}$.

Now we perform the following $\mathrm{SL}(5)$ transformation on the vertices:
\begin{equation}
    \left(\begin{array}{ccccc}
0 & 0 & 1 & 0 & 0\\
-1 & 0 & 0 & 0 & 0\\
-12 & 1 & 0 & 0 & 0\\
-26 & 2 & 2 & 1 & 0\\
-39 & 3 & 3 & 0 & 1
\end{array}\right)\cdot\left(\begin{array}{ccccccc}
0 & -1 & 0 & 0 & 0 & 80 & 42\\
0 & -12 & 1 & 0 & 0 & 468 & 246\\
1 & 0 & 0 & 0 & 0 & -1 & 0\\
-2 & -2 & -2 & 1 & 0 & -2 & -2\\
-3 & -3 & -3 & 0 & 1 & -3 & -3
\end{array}\right)=\left(\begin{array}{ccccccc}
1 & 0 & 0 & 0 & 0 & -1 & 0\\
0 & 1 & 0 & 0 & 0 & -80 & -42\\
0 & 0 & 1 & 0 & 0 & -492 & -258\\
0 & 0 & 0 & 1 & 0 & -1148 & -602\\
0 & 0 & 0 & 0 & 1 & -1722 & -903
\end{array}\right)\,.
\end{equation}
The geometry described by the polytope should be
$\mathrm{SL}(5)$ invariant. On the right hand
side, the first six columns are precisely the
toric rays of $\mathbb P^{1,80,492,1148,1722}$
mentioned in Section \ref{subsec:geometry}, and
the last column represents an exceptional divisor
resulting from resolving the singularity in this
weighted projective space. These data match those
in the KS database. Hence we have proved the
equivalence between these two descriptions of the geometry.

\bibliography{references}
\bibliographystyle{JHEP}

\end{document}